# Phenomenological Potential Wells for $C_{60}$ Shell

Running head: Potential Wells for $C_{60}$ Shell

A S Baltenkov
Institute of Ion-Plasma and Laser Technologies, Uzbek Academy of Sciences
100125, Tashkent, Uzbekistan, arkbalt@mail.ru

**Abstract**. A spatial distribution of electric charges forming a square potential well has been studied. It is shown that this potential is created by two spherical double charge layers, which is in contradiction with a real molecular structure of $C_{60}$ cage. A $C_{60}$ shell potential has been calculated in the assumption that it is formed by the averaged charge density of a neutral atom. It is shown that the phenomenological potentials simulating the $C_{60}$ shell potential should be selected from a family of potentials with a none-flat bottom. Two possible types of $C_{60}$ model potentials have been considered and their parameters have been calculated.

## 1. Introduction

A unique structure and stability of large carbon clusters such as $C_{60}$ made them a subject of numerous investigations. The density functional theory is widely used to describe electronic structure of a $C_{60}$ molecule [1-3]. Within the framework of this method the wave functions of collectivized $2s^2 2p^2$ electrons are calculated by solving the Kohn-Sham equation in the field of a positively charged shell. The latter is described usually by a uniform thin spherical shell while a deviation from the spherical symmetry due to icosahedral structure of $C_{60}$ is treated as a perturbation. The effect of very tightly bound $1s^2$ electrons forming together with carbon atom nuclei a positive spherically symmetrical potential in which the collectivized electrons are moving is described by different methods. In Ref. [4] the potential of $1s^2$ electrons was taken into account by a pseudopotential being proportional to a delta function $\delta(r-R)$ . Here $R$ is the radius of the $C_{60}$ skeleton, i.e. the distance between the center of the fullerene cage and nuclei of carbon atoms. In Refs. [5,6] another procedure was used to determine the charge density distribution of core $1s^2$-electrons. The procedure is in averaging of the $1s^2$-atomic electron density calculated in the Hartree–Fock approximation over the sphere of radius $R$. In Refs. [7,8] a constant potential $U_0$ was added to the jellium potential representing sixty $C^{4+}$ ions. The parameters of this potential, i.e. inner $r_i$ and outer $r_o$ radii and depth $U_0$, are determined by requiring both charge neutrality and obtaining the experimental value of the first $C_{60}$ ionization potential.

A square potential describing the $C_{60}$ potential well is widely used to study the photoionization of A atoms encapsulated inside the hollow interior of the $C_{60}$ fullerene cage, so called endohedral atoms A@$C_{60}$ [8]. Experimental discovery of confinement resonances in photoionization of 4*d* subshell of Xe atom in molecule $Xe@C_{60}^{+}$ [10] stimulated a number of papers (see Refs. [11-15] and references therein) with attempts to describe the experimental results by varying the square potential parameters.

The $C_{60}$ shell is formed by a superposition of the positive charge of atomic nuclei (or positive ions) and negative charge of the electron cloud. It is interesting to analyze what the spatial distribution of charges should be for the energy of electron interaction with the $C_{60}$ shell to be defined by the formula

$$U(r) = \begin{cases} -U_0, \text{ if } R\text{-}\Delta/2 \le r \le R + \Delta/2; \\ 0, \qquad \text{otherwise,} \end{cases} \tag{1}$$

here, $\Delta = r_o - r_i$ is the thickness and $U_0$ is the depth of the square potential well [11]. Sections 2 and 3 of the paper are devoted to this analysis where two model potentials associated with the potential (1) are studied. In Sec. 4 the $C_{60}$ shell potential is discussed assuming that it is created by the averaged (over the sphere of radius *R*) charge density of neutral atom. In Sec. 5 two possible types of the $C_{60}$ potential are considered and their parameters are calculated. In Sec. 6 the charge density forming a spherical potential well with zero thickness is analyzed. Section 7 gives conclusions.

## 2. Spatial distribution of charge for potential well (1)

The energy of electron interaction with the $C_{60}$ shell is represented as two Fermi steps

$$U(r) = \begin{cases} -U_0[1+\exp\{(R-r-\Delta/2)/\eta\}]^{-1} \text{ for } r \le R, \\ -U_0[1+\exp\{(r-R-\Delta/2)/\eta\}]^{-1} \text{ for } r \ge R. \end{cases} \tag{2}$$

Here, $\eta$ is the diffuseness parameter [11]. The formulas (1) and (2) coincide for the limit $\eta \to 0$. The energy of electron interaction (2) is connected with the potential of electric field $\varphi(r)$ in which the electron is moving by the relation $U(r) = -\varphi(r)$ (here it is taken into account that the electron charge is equal to -1). The atomic units (au) ($\hbar = m = |e| = 1$) are used throughout this paper. The electrostatic field potential $\varphi(r)$ is defined by the Poisson equation

$$\Delta\varphi = -4\pi\rho\,, \tag{3}$$

where $\rho(r)$ is the density of the electric charge forming the spherical symmetrical potential well (2). The radial dependence of this density is defined by the following equation

$$\frac{1}{r}\frac{d^2}{dr^2}[rU(r)] = 4\pi\rho(r)\,. \tag{4}$$

Substituting (2) in equation (4) we obtain the following expression for the charge density

$$\rho(r) = U_0\frac{1}{4\pi r\eta^2} \times \begin{cases} \dfrac{z_+}{(1+z_+)^3}[r(1-z_+) - 2\eta(1+z_+)], r \le R; \\ \dfrac{z_-}{(1+z_-)^3}[r(1-z_-) - 2\eta(1+z_-)], r \ge R. \end{cases} \tag{5}$$

Here $z_{\pm} = \exp[(\pm R \mp r - \Delta/2)/\eta]$. The calculation results for the functions *U*(*r*) and $\rho(r)$ are presented in Figs. 1 and 2. For the numerical calculations the following values of the potential well parameters were selected: $U_0 = 0.302$, *R*=6.84 and $\Delta = 1.9$ [11]. Fig. 2 demonstrates the evolution of the charge distribution with the diffuseness parameter varied within the range $0.02 \le \eta \le 0.2$. It is seen that for small $\eta$ approximately a half of the positive charges (carbon atom nuclei or positive ions $C^{4+}$) of the $C_{60}$ shell is located near the sphere with the radius $r_i$ while the rest of the positive charges is smeared over the sphere with the radius $r_o$. The positive charge of each of these spheres ($\rho > 0$) is compensated by two thin layers of negative charge ($\rho < 0$) formed by electrons. The reason for such a behavior of the function $\rho(r)$ is that the potential $U(r)$ is formed *per se* by two Heaviside step functions. Their derivatives of the first and second order are the delta functions $\delta(r-r_i)$ and $\delta(r-r_o)$ and the derivatives of these delta functions $\delta'(r-r_i)$ and $\delta'(r-r_o)$, respectively.

The $C_{60}$ shell is uncharged as a whole, therefore the total charge

$$Q=\int_0^\infty \rho(r)r^2dr=0\,, \qquad (6)$$

where the charge density $\rho(r)$ is described by formula (5).

### 3. Power-exponential potential

Considered in Ref. [16], the power exponential potential

$$U(r)=-U_0\exp[-(r-R)^p/w^p] \qquad (7)$$

allows the potential shape to be changed continuously from a Gaussian type ($p$=2) to a square-well type ($p\to\infty$). Note that the parameter $p$ is even. Let us use the potential (7) to analyze how the potential well with a flat bottom is transformed into a potential with the cusp-shaped bottom. Substituting (7) in the Poisson equation (4), we obtain the following expression for the charge density

$$\rho(r)=U_0\frac{zy}{4\pi r}\frac{p}{w^p}[2-ry\frac{p}{w^p}+r\frac{(p-1)}{(r-R)}]\,. \qquad (8)$$

where the following replacements are made

$$z=\exp[-(r-R)^p/w^p]\,; \qquad y=(r-R)^{p-1}\,. \qquad (9)$$

The calculation results for the functions $U(r)$ and $\rho(r)$ are presented in Figs. 3 and 4. For the numerical calculations the values of the potential well parameters ($w=\Delta/2$) were the same as in Sec. 1. For $p$=2 we deal with the cusp-shaped bottom. In this potential well the atomic nucleus (or the positive ions) of carbon atoms are localized near the sphere with the radius $R$. The cusp-shaped bottom of the well becomes flatter with the $p$ parameter increase. The charge density distributions in Figs. 2 and 4 are qualitatively similar and they evidence that the potential (1) can be created only in the assumption that the $C_{60}$ shell is formed by two double-charged spheres with the radii $r_i$ and $r_o$.

It is evident that such a construction of $C_{60}$ is impossible because all the fullerene nuclei are located at approximately equal distance $R$ from the center of the sphere. The fullerene nuclei can not form two concentric spheres with a gap $\Delta$ between them being almost 30% (1.9/6.84~0.28) of the skeleton radius $R$*. Therefore, the assumption that the fullerene shell potential can be represented by the formula (1) is in contradiction with the real molecular structure of the $C_{60}$ cage.

### 4. Averaged density of neutral atom charge

A more realistic form of the potential well can be obtained as a result of averaging the charge density of neutral atom over the sphere of the radius $R$. For simplicity we will assume this atom as one-electron one. According to Ref. [5], the averaged density of atomic electron is written as

$$\langle\rho_e(r)\rangle=-\frac{1}{4\pi}\int|\psi(\mathbf{r}-\mathbf{R})|^2\,d\Omega\,. \qquad (10)$$

In the center of the potential shell the electron density is finite and equal to

$$\langle\rho_e(r\to 0)\rangle=-|\psi(\mathbf{R})|^2\,. \qquad (11)$$

* In papers [7,8] this gap is more than 40% (1.5/3.54~0.42).

Since the total charge of the electronic cloud is equal to unity we have the following formula

$$4\pi\int_0^\infty \langle\rho_e(r)\rangle r^2 dr = -1 . \tag{12}$$

The potential created by this spherically symmetrical electronic cloud at the point **r** is

$$\varphi_e(r) = \int \frac{\langle\rho_e(r')\rangle}{|\mathbf{r}-\mathbf{r}'|} d\mathbf{r}' = 4\pi\left[\frac{1}{r}\int_0^r \langle\rho_e(r')\rangle r'^2 dr' + \int_r^\infty \langle\rho_e(r')\rangle r' dr'\right]. \tag{13}$$

The electronic cloud potential in the center of the sphere is

$$\varphi_e(r\to 0) = 4\pi\int_0^\infty \langle\rho_e(r')\rangle r' dr' . \tag{14}$$

Far from the sphere center for $r >> R$ we have

$$\varphi_e(r\to\infty) = 4\pi\frac{1}{r}\int_0^\infty \langle\rho_e(r')\rangle r'^2 dr' = -\frac{1}{r} . \tag{15}$$

Averaged over the sphere of the radius *R*, the density of point charge of atom nucleus is

$$\langle\rho_n(r)\rangle = \frac{1}{4\pi}\int \delta(\mathbf{r}-\mathbf{R})d\Omega = \frac{1}{4\pi}\frac{\delta(r-R)}{R^2} . \tag{16}$$

The potential created by this charge is obtained by substitution $\rho_n(r)$ into formula (13)

$$\varphi_n(r) = \begin{cases} 1/R & \text{for } r \le R, \\ 1/r & \text{for } r \ge R. \end{cases} \tag{17}$$

Thus, the potential of the electro-neutral layer formed by the neutral atom smeared over the sphere of the radius *R*

$$\varphi(r) = \varphi_n(r) + \varphi_e(r) , \tag{18}$$

is finite in the center of the fullerene sphere and tends to zero for $r >> R$ .

Let us write the wave function in the following form

$$\psi(\mathbf{r}) = \frac{z^{3/2}}{\pi^{1/2}}\exp(-zr) . \tag{19}$$

The charge densities $\langle\rho_e\rangle$ and $\langle\rho_e\rangle$ for different values of the parameter *z* are presented in Fig. 5. The averaged density of the atom nucleus charge (16) ( $\rho_n > 0$ ) is represented in this figure by the Lorentz curve

$$\langle\rho_n(r)\rangle = \frac{1}{4\pi R^2}\frac{1}{\pi}\frac{d}{(r-R)^2+d^2} . \tag{20}$$

With *z* increasing, the electron cloud with $\rho_e < 0$ is getting more and more localized at the both sides of the positively charged sphere with the radius *R*.

Calculated with the formulas (13) and (17), the potential wells $U(r) = -\varphi(r)$ formed by smeared atomic charge densities for different values of the parameter *z* are presented in Fig. 6. All these potentials have the cusp-shaped bottoms. Hence, the phenomenological potentials modeling the $C_{60}$ shell potential are to be selected from a family of curves similar to those given in Fig. 6. It is evident that the number of such potentials is unlimited. Further we will consider two model potentials for the $C_{60}$ shell.

## 5. The model potentials

An important feature of the $C_{60}$ potential well is the presence of a shallow level in it. The electron affinity of $C_{60}$ determined by UV photoelectron spectroscopy [17] is about $I = 2.7 \pm 0.1\,\text{eV}$. There is different data on the symmetry of the ground state of the ion $C_{60}^-$. The group of symmetry of fullerene $C_{60}$ causes a degenerate state of an extra electron in this ion [18,19]. In accordance with symmetry $I_h$, the ground state on the $C_{60}^-$ ion is a *p*-like state. Experiments [20] show that the electron attachment occurs in the ground state of *p*-symmetry. The fact that there is a threshold in the cross section of this process is explained by the existence of a centrifugal barrier absent for electron capture in the ion *s*-like state. In [21,22] using a different method of measuring the slow electron attachment, the threshold of this process was not found. This proves the attachment of the slow electron to the ground *s*-state. This observation is also in line with results obtained in [23] for the attachment of Rydberg electrons to $C_{60}$ demonstrating the *s*-wave capture.

Both variants of symmetry of electron attachment were considered in the models of negative ion in Refs. [24,25]. In the first variant it was assumed that the ground state of the $C_{60}^-$ ion is a *p*-like state with the binding energy $E_p$=-2.7 eV. In the second one the ground state of the $C_{60}^-$ system was the *s*-level with the binding energy $E_s$=-2.65 eV. The presence of these levels imposes limitations on the model potential parameters

Let us consider the following model potentials. The Dirac-bubble potential $U(r) = -U_0\delta(r-R)$ [24] can be considered as a limit case of the Lorentz-bubble potential

$$U(r) = -U_0 \frac{1}{\pi} \frac{d}{(r-R)^2 + d^2} \tag{21}$$

for $d \to 0$. The maximal depth of the potential (21) at $r$=$R$ is $U_{\max} = U_0 / \pi d$. The thickness of the potential well $\Delta$ at the middle of the maximal depth is $\Delta = 2d$. With $r$ increasing the potential (21) decreases as $r^{-2}$. Along with (21) we consider the Cosh-bubble potential exponentially decreases with $r$:

$$U(r) = -\frac{U_{\max}}{\cosh^n[\alpha(r-R)]}. \tag{22}$$

In further consideration we assume $n$=1. In the middle of the maximal depth of the well (22), the thickness $\Delta$ is connected with the parameter $\alpha$ as follows

$$\Delta = \frac{2}{\alpha}\ln(2+\sqrt{3}) = 2.633916/\alpha. \tag{23}$$

The parameters of these potential $U_{\max}$ and $\Delta$ should be connected with each other in such a way that in the potential wells (21) and (22) there is a *s*-like state or a *p*-like state with the specified energy $E = -I$. The parameters $U_{\max}$ and $\Delta$ are defined by solving the wave equation for the radial parts of the wave functions

$$\chi''_{nl} - \frac{l(l+1)}{r^2}\chi_{nl} + 2[E - U(r)]\chi_{nl} = 0, \tag{24}$$

where $\chi_{nl}(r) = rR_{nl}(r)$. For the fixed thickness of the potential shell $\Delta$ the parameters $U_{\max}$ should provide the solutions $\chi_{nl}(r)$ of the wave equation (22) exponentially decreasing with *r*. A set of such pairs of the parameters defines two families of the potentials $U(r)$ where the *s*-like or *p*-like ground state with the binding energy *E* exists. The equation (22) with the potentials (21) and (22) was solved by the Runge-Kutta method [26]. The thickness $\Delta$ was equal to 1, 2 and 3 au. The eigen-value in equation (24) was equal to *E*=-2.65 eV, the radius *R* to *R*=6.665 [4]. The parameter $U_{\max}$ was varied until the eigen-function $\chi_{nl}(r)$ vanishes at large distances from the center of the $C_{60}$ cage. The calculation results for these functions for the orbital moment *l*=0 (*s*-like ground state) and for *l*=1 (*p*-like ground state) are presented in Fig. 7. For comparison, the wave functions calculated in [24,25] with the Dirac-bubble potential ($\Delta = 0$) are given in the same figures. With the rise in the parameter $\Delta$ the cusp-behavior of the wave function for zero-thickness changes to more smoothly behavior of $R_{nl}(r) = \chi_{nl}(r)/r$ near the point $r \approx R$. According to Fig. 7, the shape of the wave functions depends comparatively weakly on the shape of the potential wells in which the electron is localized. The potentials with the calculated parameters $\Delta$ and $U_{\max}$ are shown in Fig. 8. As expected, the Cosh-bubble potential is localized in narrower (as compared to the Lorentz type potential) region near the sphere of the radius *R*.

**6. The Dirac-bubble potential**

In Sec. 2 we discussed the connection of the charge density for potential (1) with delta functions $\delta(r-r_i)$ and $\delta(r-r_o)$ and with derivatives of these functions $\delta'(r-r_i)$ and $\delta'(r-r_o)$. Now let us analyze the charge density corresponding to the delta-bubble potential (21). Substituting the potential (21) in the Poisson equation (3) we obtain the following expression for the density of charges forming the Dirac-bubble potential

$$\rho(r) = U_0 \frac{d}{2\pi^2 r[z^2+d^2]^3}[2z(z^2+d^2)+r(d^2-3z^2)], \qquad (25)$$

where $z = r - R$. Fig. 9 demonstrates the evolution of the charge density for $0.1 \le d \le 0.5$. Here, as in Sec. 2, the following values of the parameters were used: $U_0 = 0.302$ and *R*=6.84. The charge density, as in Fig. 5, forms a three-layer sandwich: the middle layer is positively charged (atomic nuclei or positive ions $C^{4+}$) and the outer layers are negatively charged (electron clouds). In the limit $d \to 0$ all these spherical layers have zero thickness and radius equal to the skeleton radius *R,* which is not in contradiction with the data on molecular structure of the $C_{60}$ cage. The reason for such a behavior of the function $\rho(r)$ is that charge density, in this case, is defined by the second derivative $\delta''(r-R)$ rather than the first derivatives of the delta-functions as in Sec. 2.

**7. Conclusions**

It has been shown that the potential wells with a flat bottom should be eliminated from the family of phenomenological potentials when considering the $C_{60}$ shell. According to the calculations in Sec. 2 and 3, these wells should be formed only by a superposition of the electric charges with the spatial distributions being in contradiction with the data on molecular structure of the $C_{60}$ cage. The two model potentials for the $C_{60}$ shell with a cusp-shaped bottom have been considered. The parameters of these potentials providing existence of bound states of negative ion $C_{60}^{-}$ have been calculated. It should be noted that representation of the fullerene shell potential as a model function $U(r)$ is essential idealization of a real potential of the $C_{60}$ cage and therefore cannot describe all features of either $C_{60}$ molecule or endohedral atom A@$C_{60}$.

**Acknowledgments**

The author is very grateful to Dr. I. Bitenskiy for useful comments. This work was supported by the Uzbek Foundation Award Ф2-ФА-Ф164.

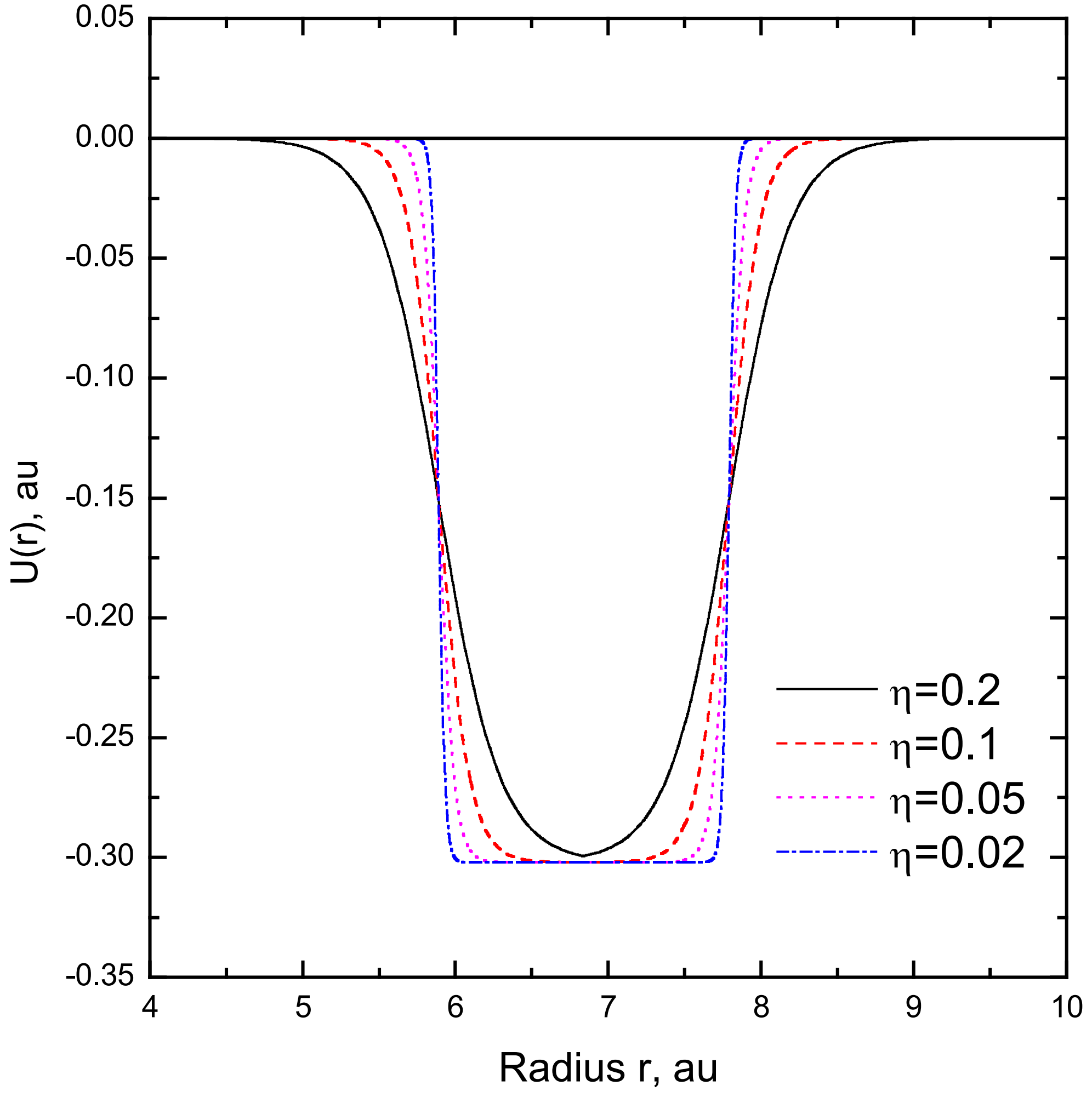

Fig. 1. The potential wells Eq. (2) for different diffuseness parameters

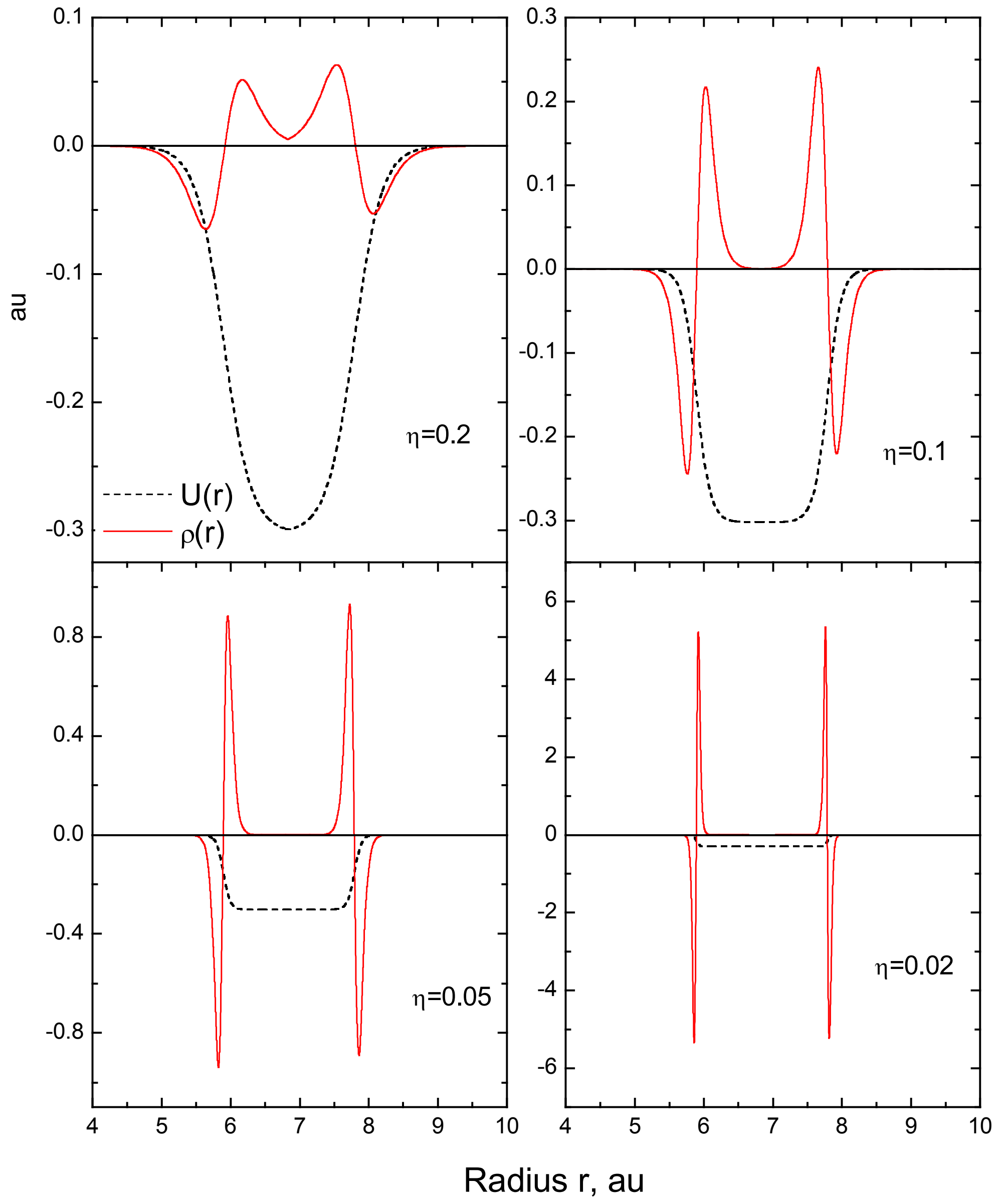

Fig. 2. The potentials $U(r)$ and charge densities $\rho(r)$ for different diffuseness parameters $\eta$

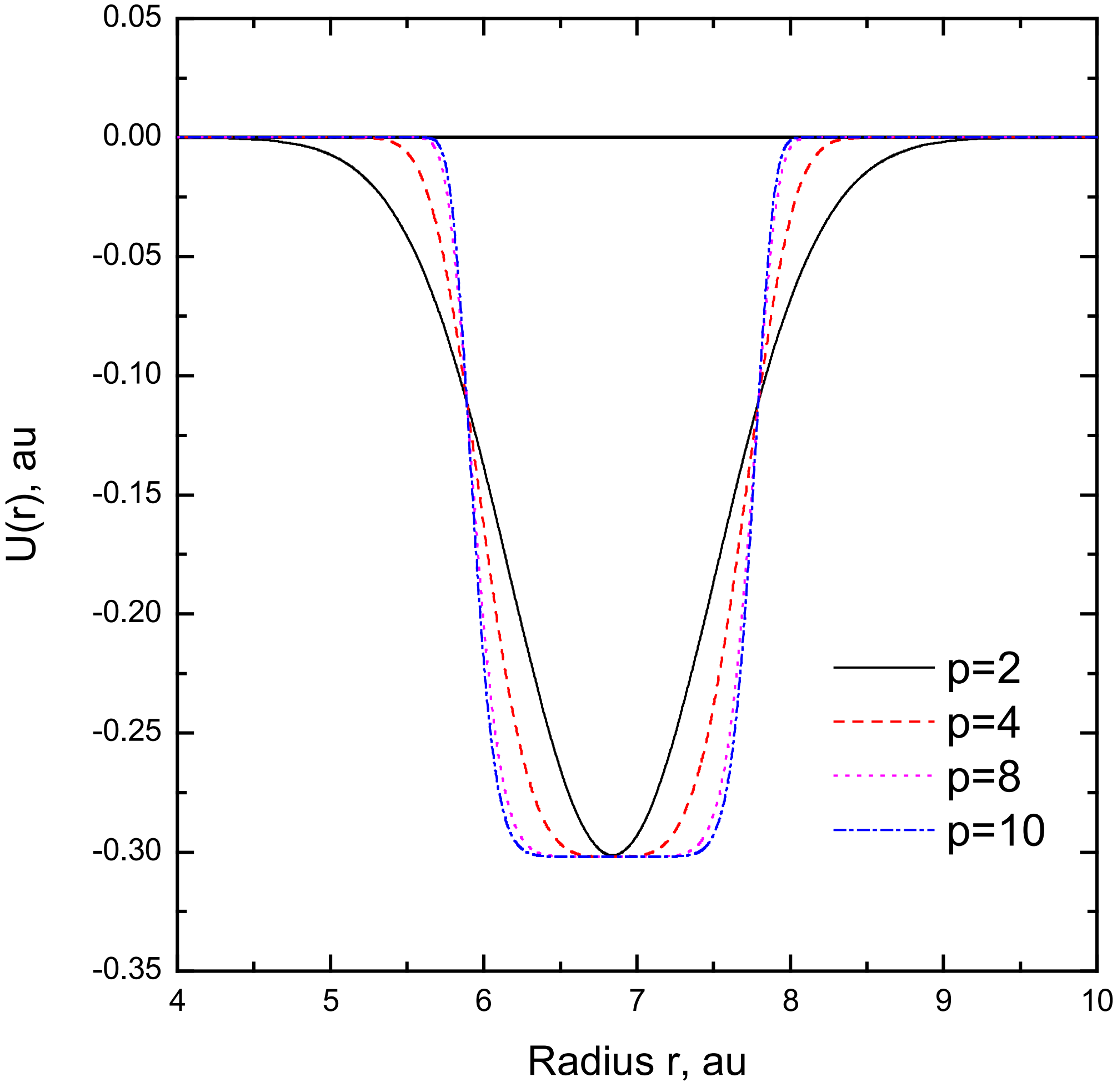

Fig. 3. The potential wells Eq. (7) for different parameters *p*

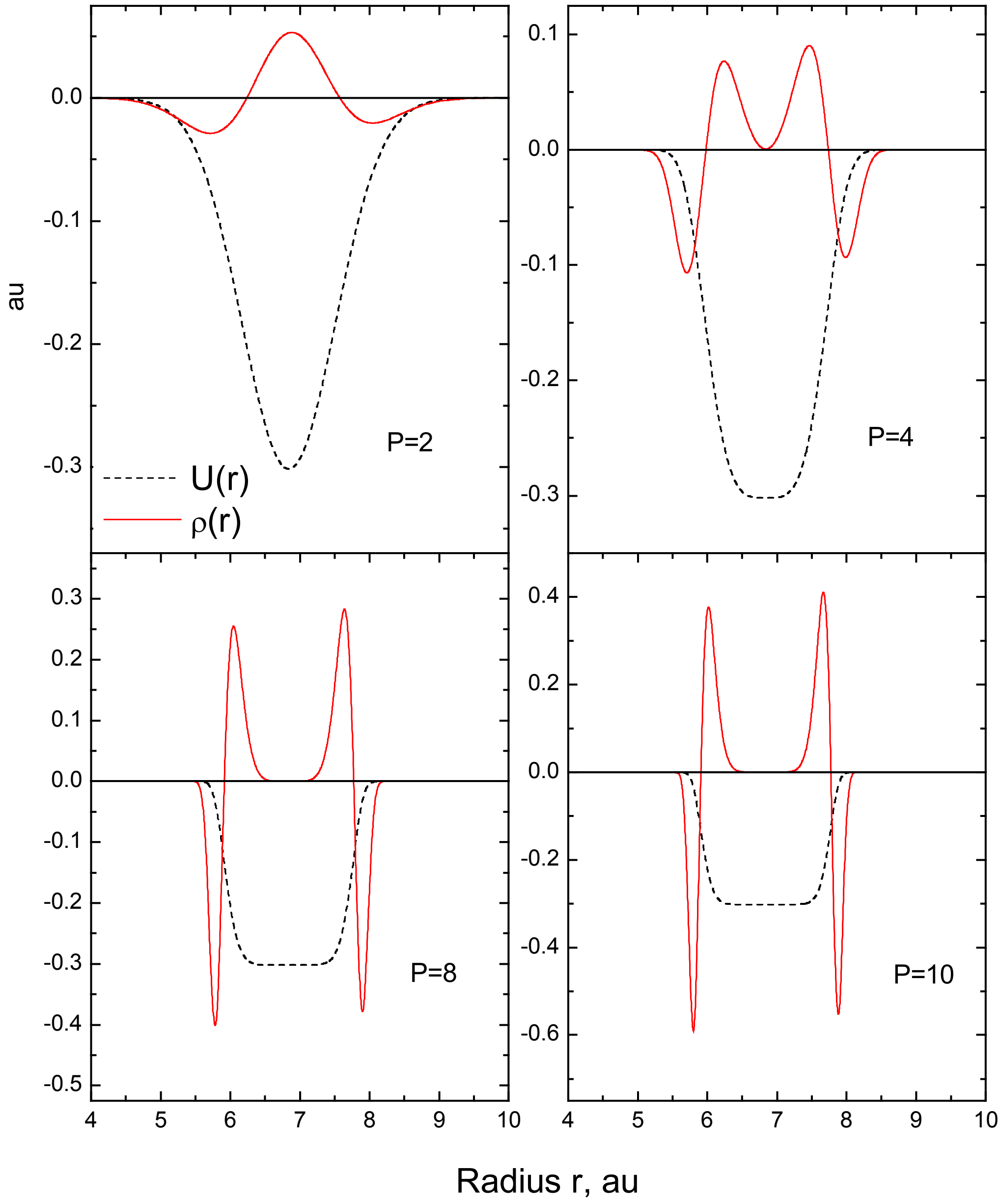

Fig. 4. The potentials $U(r)$ (7) and charge densities $\rho(r)$ (8) for different parameters $p$

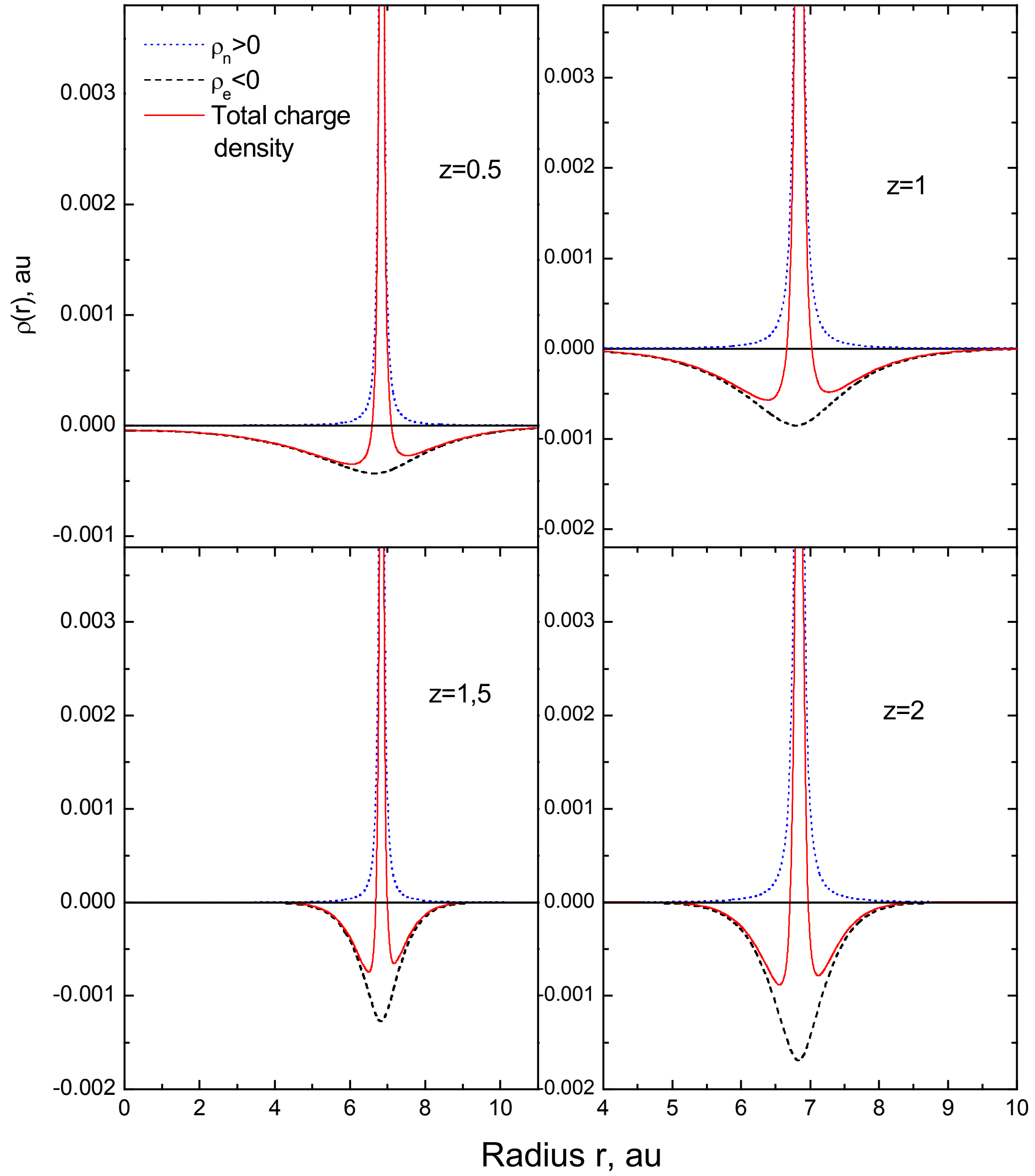

Fig. 5. The charge densities (18) for different parameters *z*

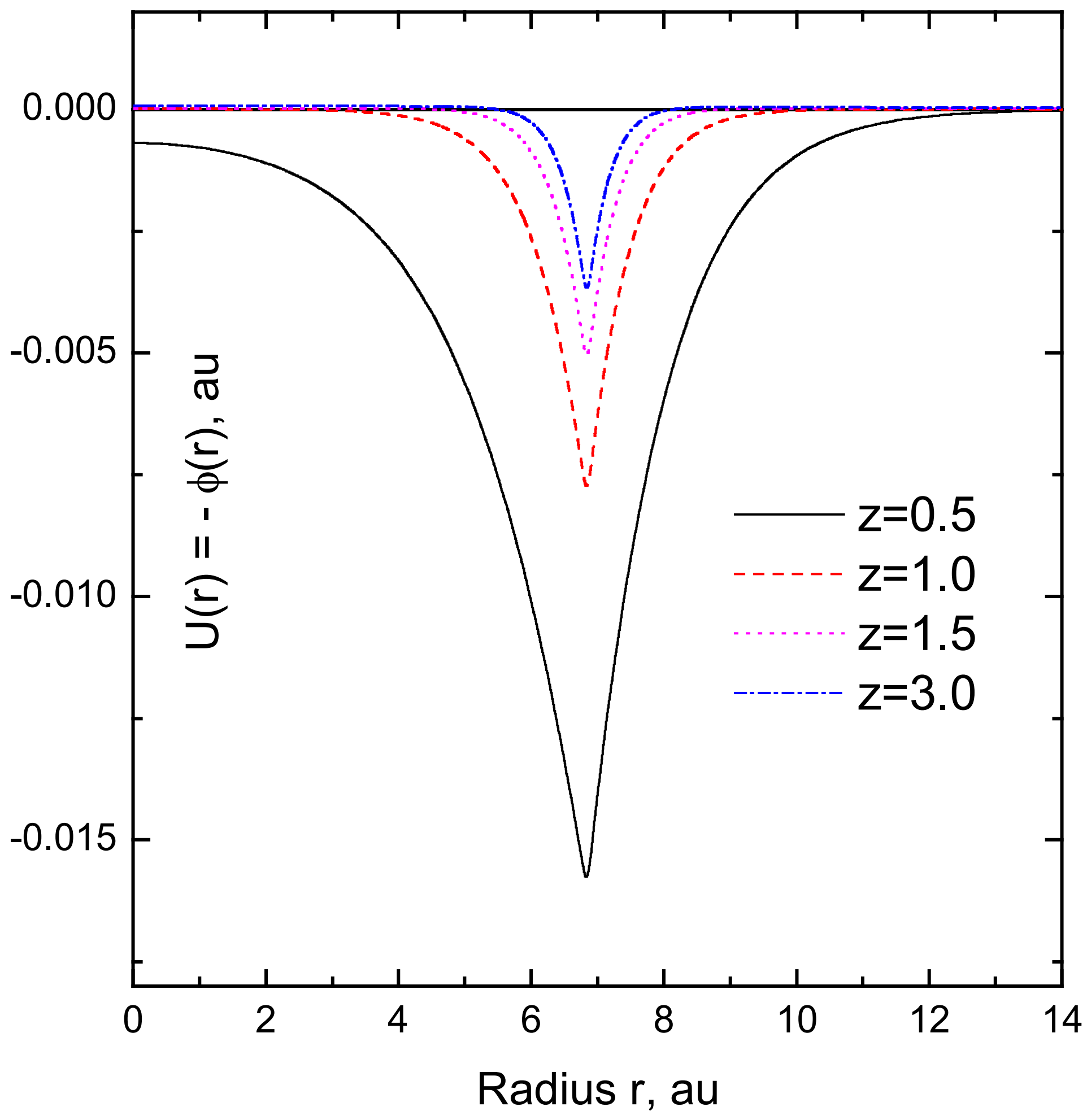

Fig. 6. The potential wells Eqs. (13) and (17) with different $z$

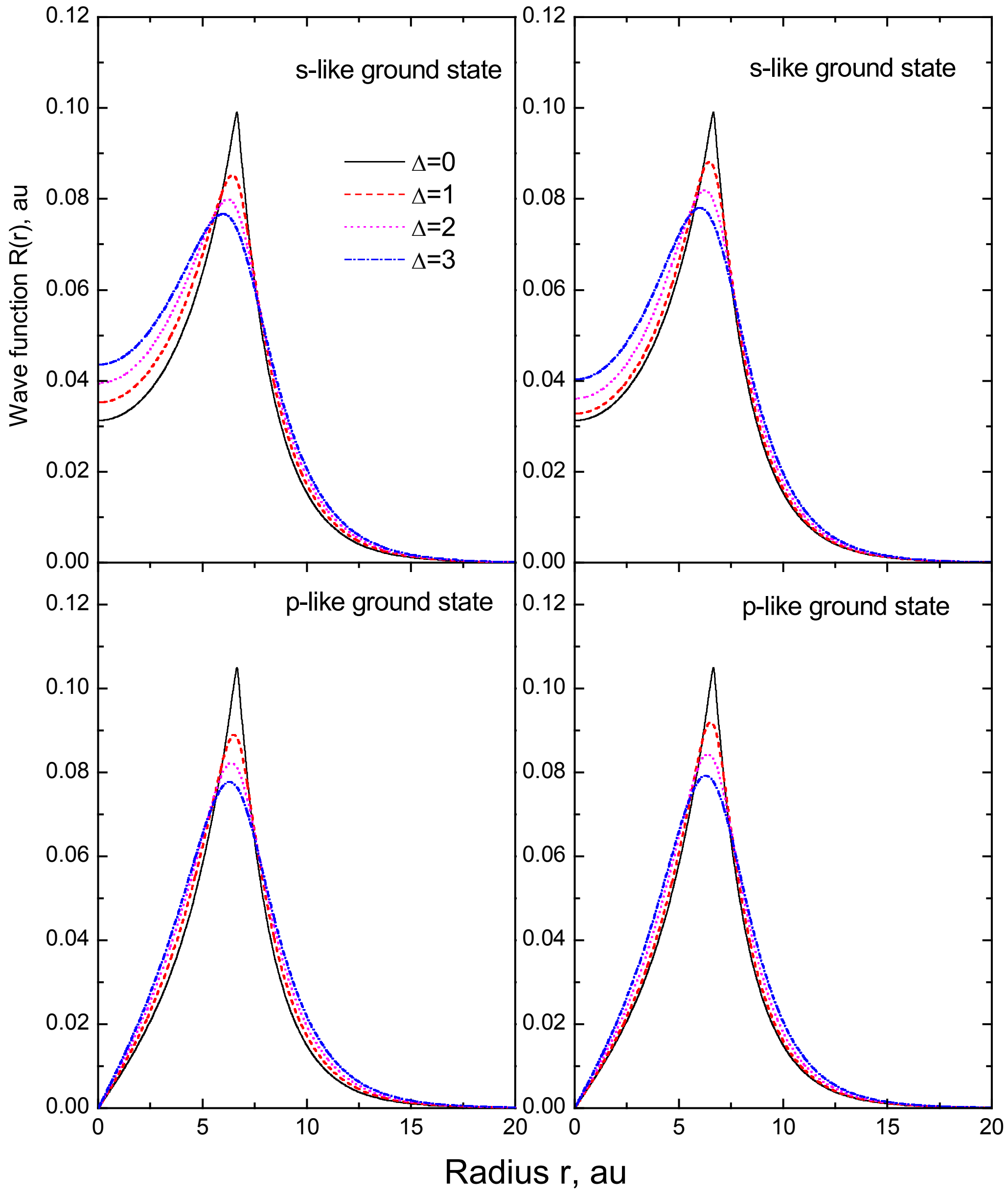

Fig. 7. The wave functions $R(r)$. The left panel is the Lorentz-bubble potential Eq. (21). The right panel is the Cosh-bubble potential Eq. (22).

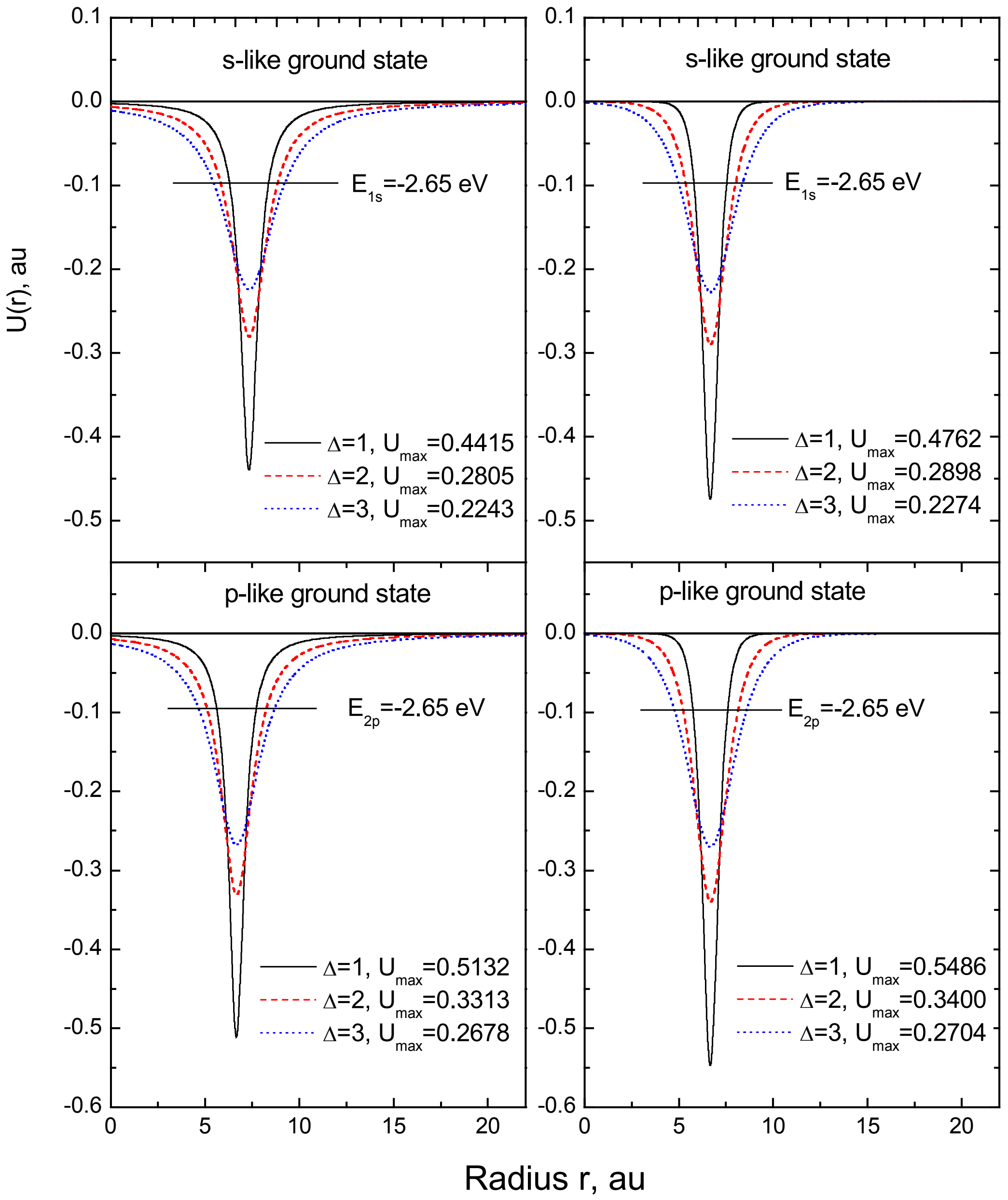

Fig. 8. The model potentials and their parameters. The left panel is the Lorentz-bubble potential Eq. (21). The right panel is the Cosh-bubble potential Eq. (22).

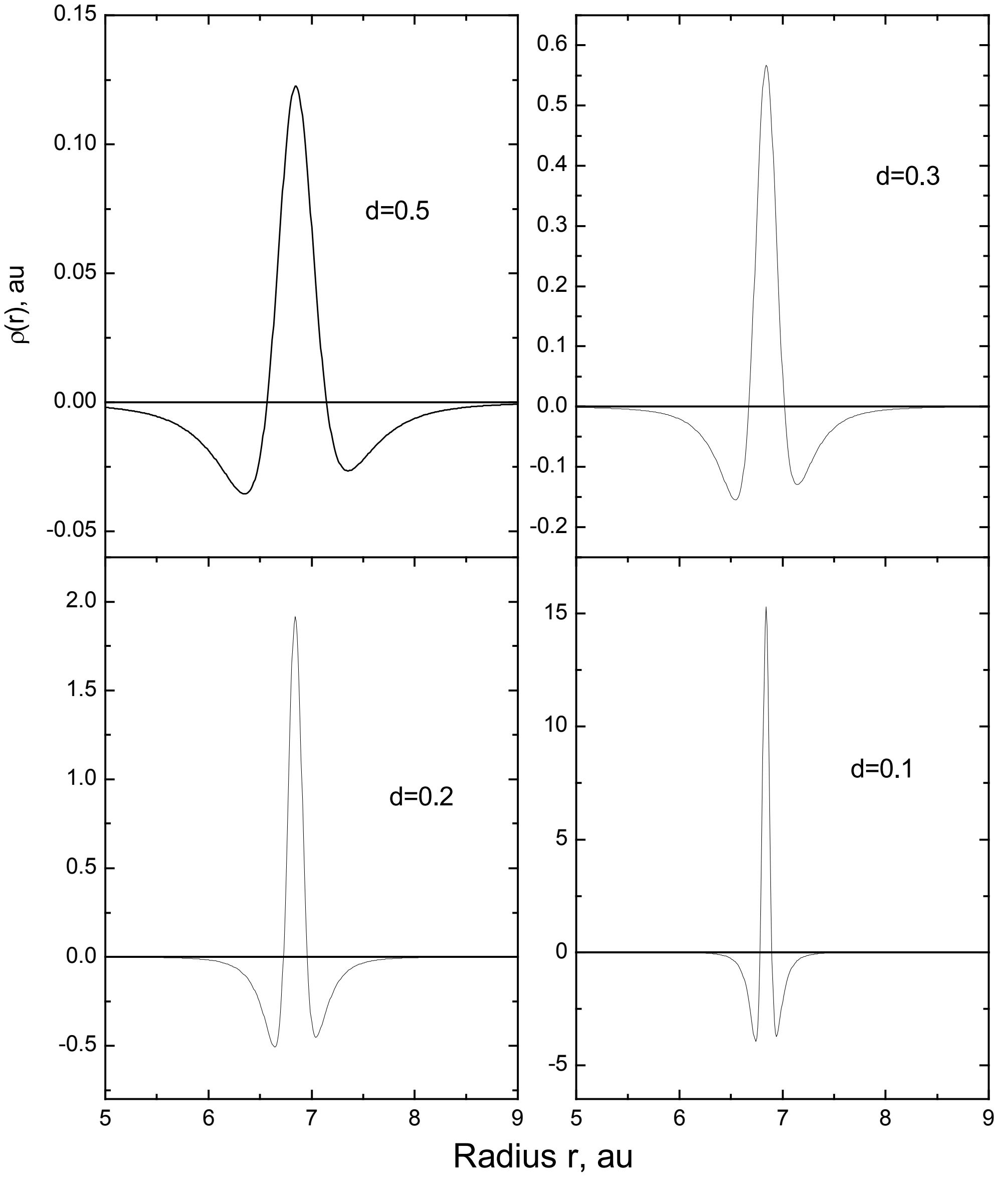

Fig. 9. The charge densities forming the Dirac-bubble potential Eq. (21) for different half-widths $d = \Delta / 2$